\documentclass[apj,numberedappendix]{emulateapj}
\usepackage[]{natbib}
\usepackage{graphicx,psfig,ulem,epsfig}
\usepackage{epstopdf}
\usepackage{txfonts}
\usepackage{soul,color}
\usepackage{longtable}

\usepackage{color}

\newcommand{\simgt}{\lower.5ex\hbox{$\; \buildrel > \over \sim \;$}}
\newcommand{\simlt}{\lower.5ex\hbox{$\; \buildrel < \over \sim \;$}}

\shorttitle{XMM-\textit{Newton}/SDSS: star formation efficiency and 
$\Omega_{\rm m}$ constraints} \shortauthors{Lagan\'{a} et al.}

\begin{document}

\DeclareGraphicsExtensions{.pdf,.png,.gif,.jpg}

\title{XMM-\textit{Newton}/SDSS: star formation efficiency in galaxy clusters and
constraints on the matter density parameter}

\author{Tatiana F. Lagan\'{a}\altaffilmark{1,2}} 

\author{Yu-Ying Zhang\altaffilmark{2,3}}

\author{Thomas H. Reiprich\altaffilmark{2}}

\author{Peter Schneider\altaffilmark{2}}

\altaffiltext{1}{Universidade de S\~ao Paulo, Instituto de Astronomia, 
Geof\'isica e Ci\^encias Atmosf\'ericas, Departamento de Astronomia, Rua do Mat\~ao 1226,
Cidade Universit\'aria, CEP:05508-090, S\~ao Paulo, SP, Brasil.}
\altaffiltext{2}{Argelander-Institut f\"ur Astronomie, Universit\"at Bonn, Auf
  dem H\"ugel 71, 53121 Bonn, Germany}
\altaffiltext{3}{National Astronomical Observatories, Chinese Academy of Sciences, Beijing, 100012, China}

\begin{abstract}

It is believed that the global baryon content of clusters
of galaxies is representative of the matter distribution of the universe, and 
can, therefore, be used to reliably determine the matter density parameter $\Omega_{\rm m}$.
This assumption is challenged by the growing evidence from optical 
and X-ray observations that the total baryon mass 
fraction increases towards rich clusters. 
In this context, we investigate the dependence of stellar, and total
baryon mass fractions as a function of mass.
To do so, we used a subsample of nineteen clusters extracted from the X-ray flux limited 
sample HIFLUGCS that have available DR-7 \textit {Sloan Digital Sky Survey} (SDSS) data. 
From the optical analysis we
derived the stellar masses. Using XMM-\textit{Newton} we derived the 
gas masses. Then, adopting a scaling relation we estimate the total masses.
Adding the gas and the stellar mass fractions 
we obtain the total baryonic content that we find to 
increase with cluster mass, reaching  \textit {7-year Wilkinson Microwave Anisotropy Probe}
(WMAP-7) prediction for clusters with 
$M_{500} = 1.6 \times 10^{15} M_{\odot}$. 
We observe a decrease of the stellar mass fraction (from 4.5\% to $\sim$1.0\%) with
increasing total mass where our findings for the stellar mass fraction agree with previous studies. 
This result suggests 
a difference in the number of stars formed per unit of halo 
mass, though with a large scatter 
for low-mass systems. That is, the efficiency of star formation varies on cluster 
scale that lower mass systems are likely to have higher star formation efficiencies.
It follows immediately that the dependence of the stellar mass
fraction on total mass results
in an increase of the mass-to-light ratio from lower to higher mass systems.
We also discuss the consequences of these results in the context of determining 
the cosmic matter density parameter $\Omega_{\rm m}$.

\end{abstract}

\keywords{cosmology: observations}

\section{Introduction}
\label{intro}
One of the key goals of observational cosmology is to
determine the matter density of the universe ($\Omega_{\rm m}$).
One of the classic methods of inferring $\Omega_{\rm m}$ is to 
use the \citet{oort58} technique adopting the
mass-to-light ratios of galaxy clusters. Independently, one can 
assume that the ratio of baryonic-to-total mass of very massive galaxy clusters should closely match
the ratio of the cosmological parameters, and thus
$\Omega_{\rm b}/\Omega_{\rm m} \sim M_{\rm b}/M_{\rm tot}$ 
\citep{white93,evrard97,ettori03b,allen08}. 

In these two cases, it is widely believed that the 
global baryon content and the mass-to-light ratio 
of galaxy clusters are fairly representative of the matter distribution of the 
universe, and can therefore be used, to reliably determine the cosmic 
matter density parameter.
Contrary to expectations, this fundamental assumption is 
challenged by the growing evidence from optical and X-ray observations that the total baryon mass fraction 
\citep{david90,lms03,gonzales07,giodini09} and the mass-to-light ratio 
\citep[e.g.,][]{adami98,girardi00,bahcall02,rines04,muzzin07}
increase towards rich clusters for a broad mass range of systems. 
Also the comparison between recent direct measurements of the
baryon mass content of galaxy clusters and the prediction from the WMAP-7 \citep{jarosik11}
presents controversial mass dependence of missing baryons \citep{andreon10,
giodini09,gonz07}.

Observational difficulties prevented recent studies from building a large 
sample of galaxy clusters where both masses and luminosities are 
homogeneously computed, what consequently affect the 
reliability of $\Omega_{\rm m}$ estimates \citep{girardi00}. From the optical point of view, 
the determination of cluster galaxy luminosities is limited by uncertainties
related to corrections for Galactic extinction and background galaxy contamination, 
extrapolation of the luminosity function towards the faint-end, and the completeness 
of the sample. Also, various methods can be applied to estimate cluster mass.
The total masses can be inferred from either X-ray or optical data, under the assumption of 
hydrostatic equilibrium. Estimates based on gravitational lensing do not require 
assumptions about the dynamical state of the cluster, but it encompasses substructures 
along the line of sight. Recent numerical simulations \citep{meneguetti10} claim that 
combining weak and strong lensing data, the projected masses within 
$r_{200}$\footnote[1]{the radius at which the mean mass density is 200 times the 
critical density at the cluster redshift.}  
can be 
constrained with a precision of $\sim 10\%$. However, deprojection of lensing masses 
increases the scatter around the true masses by more than a factor of two because of 
cluster triaxiality. X-ray mass measurements have much smaller scatter (about a factor of 
two less than the lensing masses), but they are generally biased 
toward low values between 5$\%$ and 10$\%$.

The baryons in galaxy
clusters consist of stars in cluster galaxies, intra-cluster light
\citep[ICL, stars that are not bounded to the cluster galaxies;
e.g.,][]{gonzales07,kb07}, and the hot intra-cluster gas
\citep[ICG,][]{sarazin77}. The baryonic mass of a galaxy cluster is 
dominated by the ICG, the mass of which exceeds the mass of the former
two components by a factor of $\sim 6$. Thus, to reliably compute this cosmological 
parameter from the baryonic-to-total mass ratio one should address all baryonic 
components and not only the gas mass.
In the past years, measuring individual baryon budget in galaxy
clusters became possible with various high-quality optical imaging
data \citep[e.g.,][]{david90,lms03,roussel00,gonz07,laga08}. The
combination of the X-ray measured gas mass fraction ($f_{\rm gas}=
M_{\rm gas}/M_{\rm tot}$, ratio of the gas-to-total mass) with the
optically measured stellar mass fraction ($f_{\star}=M_{\star}/M_{\rm
tot}$, ratio of the stellar-to-total mass) can thus better constrain
the total baryon mass fraction ($f_{\rm b} = f_{\star} + f_{\rm gas}$). 
Notably, these observational measurements are essential to improve our 
understanding on the distributions of each of these components 
on cluster scales. Besides, studies of structure 
formation require constraints on
the baryonic components in galaxy clusters to understand star
formation and metal enrichment processes. 

In this context, HIFLUGCS \citep{reip02,hudson10} provides 
an X-ray flux-limited sample of
64 nearby ($z \le 0.1$) clusters selected from the \textit{ROSAT} All-Sky
Survey (RASS). Nineteen clusters in the HIFLUGCS sample have available 
\textit{Seventh Data Release of Sloan Digital Sky Survey} data \citep[DR-7 SDSS;]{aba09}.
It should be mentioned that the flux-limited samples 
might be biased towards more luminous clusters, leading to a
larger fraction of cool-core clusters for lower mass systems
and also they are not necessarily unbiased  with respect to morphology 
\citep{reip06}. But the advantage of working with flux-limited 
samples is that the bias can 
be determined \citep[e.g.,][]{ikebe02,stanek06,eckert11}.

The investigation of the total baryon mass fraction and 
mass-to-light ratio dependence on total cluster mass 
is one of the key goals to understand better the dynamical history of galaxy clusters
and also the use of these systems to determine the cosmological parameter $\Omega_{\rm m}$.
Thus, in this paper, we analyze
the X-ray XMM-\textit{Newton} and optical DR-7 SDSS imaging data for 
the nineteen above mentioned clusters. 
It is important to stress that both luminosities and 
total masses are determined in a homogeneous way.
Using the X-ray gas mass as a proxy to determine the 
total mass seems to be the more suitable technique 
since it is known that gas mass may likely be unbiased mass proxy 
\citep{Okabe10}.

The paper is organized as follows. In Sect.~\ref{data} we describe the
sample selection criteria providing X-ray and optical details
separately. We discuss our results in Sect.~\ref{res}, and summarize
our findings in Sect.~\ref{conc}. The scaling relation used to determine the total mass
from the gas mass in given in Appendix~\ref{mmmg}, extra information about the clusters 
(such as the color-magnitude diagrams, CMD, and luminosity function fits) 
is given in Appendix~\ref{ap}, and a detailed discussion about the main
systematic uncertainties involved to derive the stellar masses 
are presented in Appendix~\ref{opt_systeffec}. Throughout this paper we assume 
$\Omega_{\rm m}=0.3$, $\Omega_\Lambda=0.7$ and, $H_{0} = 70
\rm ~km~\rm s^{-1}~\rm Mpc^{-1}$. Confidence
intervals correspond to the 68\% confidence level. 
We adopted these values to be consistent with the X-ray analysis presented in \citet{Zhang11}.
Even so, it should be noted that the the WMAP-7 parameters for the standard $\Lambda$CDM model are
$\Omega_{\rm m}=0.267 \pm 0.0288$, $\Omega_\Lambda=0.734 \pm 0.029$, and $H_{0} = (71 \pm 2.5)
\rm~ km~\rm s^{-1}~\rm Mpc^{-1}$.

\section{Sample and Data Analysis}
\label{data}
\begin{figure}
\centering
\includegraphics[angle=270,width=9.cm]{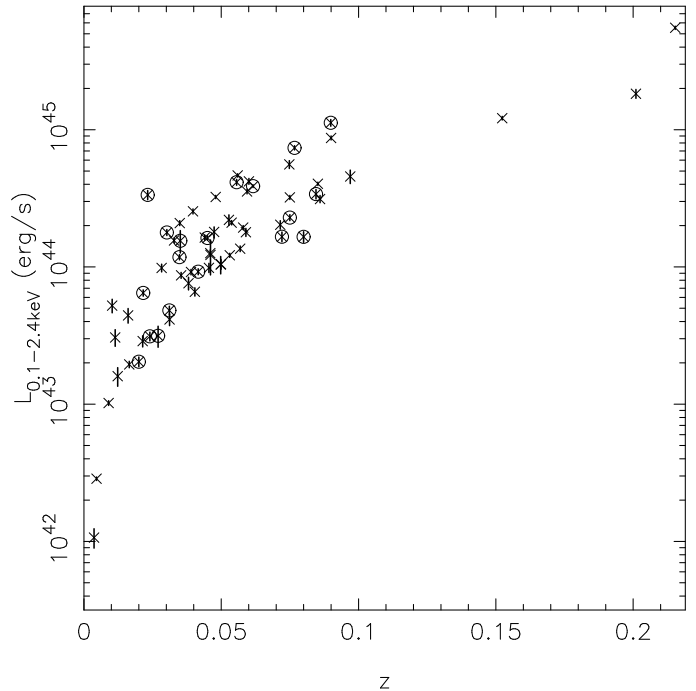}
\caption{XMM-\textit{Newton} measured X-ray bolometric luminosity within 
the cluster radius $r_{500}$ in \citet{Zhang11} vs. 
redshift for the HIFLUGCS sample. The 19 clusters with 
SDSS imaging data are highlighted by circles.}
\label{Lxz}
\end{figure}

To constrain the mass dependence of the star formation efficiency (SFE) and
baryon mass fraction, it is important to use
representative samples. The HIFLUGCS \citep{reip02} provides an X-ray
flux-limited sample of 64 nearby ($z \le 0.1$) clusters selected from the
RASS. Nineteen clusters in the HIFLUGCS have been observed in the DR-7 SDSS
\citep{aba09} and are unbiased with respect to the original HIFLUGCS
sample, having the same nature of flux selection, as can be seen in Fig.~\ref{Lxz}.

\subsection{X-ray data}
\label{xray}

Detailed XMM-\textit{Newton} data reduction can be found in 
\citet{Zhang09,Zhang11} including data screening, pointing source
subtraction, background modeling, and spectral imaging analysis.

For sample studies, one has to derive all quantities consistently
within a characteristic cluster radius.
We define them in terms of $r_{500}$, the radius at which the 
mean mass density within the cluster is 500 times the 
critical density at the cluster redshift.

We can measure $r_{500}$ from the X-ray measured mass distribution
derived under the assumption of hydrostatic equilibrium \citep[HE; e.g.][]{Zhang09}. 
However, the present sample contains relaxed and non relaxed clusters where the
HE may not be valid \citep[e.g., ][]{Zhang08,Zhang10}. 
Thus, we used a scaling relation between the gas and the total mass.
To do so,  we used 41 dynamically relaxed groups and clusters collected from 
\citet{Vik06}, \citet{Arnaud07}, \citet{Bohringer07}, and \citet{Sun09}.
We then fit a power-law to the data points and we used this fit to compute total masses 
(see Appendix~\ref{mmmg} for more details): 
\begin{eqnarray}
\label{mgmtot}
\log \bigg(\frac{M_{500} \rm E(z)}{10^{14} M_{\odot}}\bigg) & = & (0.891 \pm 0.017) +\nonumber \\
& + & \log \bigg(\frac{M_{\rm gas}\rm E(z)}{10^{14} M_{\odot}}\bigg) 
\times (0.827 \pm 0.022),
\end{eqnarray}
where $E(z)=[\Omega_{m}(1+z)^{3} + (1-\Omega_{m}-\Omega_{\Lambda})(1+z)^{2}+\Omega_{\Lambda}]^{1/2}$,
describing the redshift evolution of the Hubble parameter.

We note that the following scaling relation is based on four different samples of groups and clusters
what should reduce the potential bias produced by sample selection. Besides,
these 41 systems used to construct our scaling relation are all relaxed systems, and thus
the assumption of hydrostatic equilibrium should be valid for these obejects and thus minimize the
bias in total mass estimates for the clusters in our sample.
In the way the total mass is computed, 
$\log M_{\rm tot}=\rm A + B\times \log M_{\rm gas}$, when the slope parameter 
(B in this case) is positive and lower than 1, the gas mass fraction encreases with total mass 
(see Appendix~\ref{mmmg} for more details).

\subsection{SDSS data}
\label{photodata}
We selected all HIFLUGCS clusters with DR-7 SDSS \citep{aba09} data available.
The DR-7 includes more accurate results than previous data releases, 
updated and more accurate magnitudes for all sources in the \textit{PHOTO} 
catalog, and better accounting for sky background
subtraction. Specifically, we use the \textit{dered} table of magnitudes. 
Three clusters (A119, NGC507 and, A2063) are located on the border of SDSS field of view, so that
we had to exclude them. We ended up with 19 clusters spanning the redshift 
range of $0.02 < z < 0.1$ 
listed in Table~\ref{tab_sample}. 
 
We note that there are 2MASS \citep{2mass06} observations for the 
HIFLUGCS clusters in which the K-band luminosity
data have also been used to derive stellar masses. However, SDSS data allow us to make
a cleaner selection of the cluster galaxies and to avoid fore- and background contaminations, which are
important to derive reliable luminosity functions (LF) and thus the stellar mass for our purpose.

\begin{table*}
\small
\begin{center}
\caption{HIFLUGCS clusters with DR-7 SDSS available data}
\begin{tabular}{ccr@{.}lccr@{.}lr@{.}l}
\hline 
Name & R.A. & \multicolumn{2}{c}{DEC} &  redshift & $r_{500}$ (Mpc)& \multicolumn{2}{c}{$M_{\rm gas} (10^{13} M_{\odot})$} & \multicolumn{2}{c}{$M_{500} (10^{14} M_{\odot})$}\\
\hline
A85	& 10.46	 &$-$9&30 &0.0556  &1.216 $\pm$ 0.058  & 7&46 $\pm$ 0.34 & 5&68 $\pm$ 0.81\\
A400 	& 44.42	 &6   &02 &0.0240  &0.712 $\pm$ 0.034  & 1&00 $\pm$ 0.04 & 1&07 $\pm$ 0.15\\
IIIZw54 & 55.32	 &15  &40 &0.0311  &0.731 $\pm$ 0.035  & 1&10 $\pm$ 0.21 & 1&18 $\pm$ 0.17\\
A1367 	& 176.01 &19  &95 &0.0216  &0.893 $\pm$ 0.043  & 2&24 $\pm$ 0.07 & 2&11 $\pm$ 0.30\\
MKW4 	& 181.11 &1   &90 &0.0200  &0.580 $\pm$ 0.028  & 0&47 $\pm$ 0.02 & 0&58 $\pm$ 0.08\\
ZwCl1215& 184.42 &3   &66 &0.0750  &1.098 $\pm$ 0.052  & 5&40 $\pm$ 0.29 & 4&34 $\pm$ 0.62\\
A1650 	& 194.67 &$-$1&76 &0.0845  &1.087 $\pm$ 0.052  & 5&36 $\pm$ 0.90 & 4&28 $\pm$ 0.61\\
Coma	& 194.90 &27  &96 &0.0232  &1.278 $\pm$ 0.061  & 8&18 $\pm$ 0.62 & 6&21 $\pm$ 0.89\\
A1795	& 207.22 &26  &59 &0.0616  &1.118 $\pm$ 0.053  & 5&57 $\pm$ 0.14 & 4&46 $\pm$ 0.63\\
MKW8   	& 220.18 &3   &46 &0.0270  &0.715 $\pm$ 0.034  & 1&02 $\pm$ 0.14 & 1&10 $\pm$ 0.16\\
A2029   & 227.73 &5   &74 &0.0767  &1.275 $\pm$ 0.061  & 9&17 $\pm$ 0.33 & 6&81 $\pm$ 0.97\\
A2052	& 229.19 &7   &02 &0.0348  &0.875 $\pm$ 0.042  & 2&13 $\pm$ 0.11 & 2&03 $\pm$ 0.29\\
MKW3S	& 230.47 &7   &71 &0.0450  &0.905 $\pm$ 0.043  & 2&44 $\pm$ 0.10 & 2&29 $\pm$ 0.33\\
A2065	& 230.60 &27  &71 &0.0721  &1.008 $\pm$ 0.048  & 4&12 $\pm$ 1.29 & 3&35 $\pm$ 0.48\\
A2142	& 239.58 &27  &23 &0.0899  &1.449 $\pm$ 0.069  & 15&10 $\pm$ 0.88 & 10&26 $\pm$ 1.47\\
A2147 	& 240.57 &15  &97 &0.0351  &1.064 $\pm$ 0.050  & 4&35 $\pm$ 0.43 & 3&63 $\pm$ 0.52\\
A2199 	& 247.16 &39  &55 &0.0302  &0.957 $\pm$ 0.045  & 2&90 $\pm$ 0.23 & 2&64 $\pm$ 0.38\\
A2255	& 258.12 &64  &07 &0.0800  &1.072 $\pm$ 0.051  & 5&13 $\pm$ 0.20 & 4&08 $\pm$ 0.58\\
A2589	& 350.99 &16  &78 &0.0416  &0.848 $\pm$ 0.040  & 1&93 $\pm$ 0.12 & 1&88 $\pm$ 0.27\\
\hline
\end{tabular}
\label{tab_sample}
\end{center}
\end{table*}

In order to obtain the stellar mass of galaxies we first constructed 
color-magnitude diagrams (CMD; ($g- i) ~ \times ~ i$) to statistically select cluster 
members within $r_{500}$.
In this paper, we fit iteratively the red-sequence (RS), first identified in the 
$0.7 < (g-i) < 1.6$ color interval and $i < 18$ magnitude limit. 
This interval was the one that best suited the majority of clusters in our sample.
For four clusters (IIIZw54, A1367, MKW4, and A400) we had to extend
the magnitude limit up to $i < 16$ to better define the slope of the RS. 
In this way, we select just the brighter and redder galaxies, what enables 
us to first identify the red-sequence on the CMD,
fitting a linear relation to these galaxies. Then, to constrain the position and the slope of the 
RS on the CMD, we use the whole color and magnitude interval to re-do the linear fit until
it converges. 
For red galaxies, we adopted $\pm$0.3 mag width from the red-sequence best fit.
Then, the blue galaxies were considered as the galaxies that lay below the lower limit 
of the red-sequence (see Figure~\ref{cmr} in Appendix~\ref{ap}).
To take into account fore/background contamination we computed the number 
of galaxies per 0.5 bin of magnitude in an annular region beyond
$8 \times r_{500}$ from the cluster center.
We then fit the galaxy number counts in the $20 < i < 16$ 
range of magnitude using a power-law. 
The final cluster magnitude number counts were obtained by subtracting 
the background fit from the galaxy counts.

For the selected blue and red members, we computed the luminosity function
in the \textit{i}-band (see Fig.~\ref{cmr} in Appendix~\ref{ap}), 
performing an analytical fit using two
 Schechter functions \citep{schech76} with the usual parameters
$\alpha$ (faint-end slope), $\Phi^{\ast}$ (normalization) and 
$M^{\ast}$ (characteristic magnitude). For all clusters in our sample but ZwCl1215, A1650, and A2255 
we used two Schechter functions to take into account both bright- and faint-end contribution. 
With the improvement of optical instruments, in the past years much work has been done to 
constrain the faint-population (-18 $ < M_{\rm i} <$ -12) of nearby galaxy clusters \citep[e.g.,][]{penny08,rines08,boue08}.
Since the LF of our clusters showed a steep population faintward we fit 2 Schechter functions to
describe the overall behavior of galaxies. The covariance between the characteristic magnitude
and the power-law index is taken into account (see Appendix~\ref{ap} for more details about LF fits).

We considered \textit{K}-correction ($K(z)$) and evolutionary correction ($\epsilon(z)$) supplied 
by \citet{poggi97} to correct $M^{\ast}$ values.
To derive the total luminosity we integrated the luminosity function assuming a lower luminosity limit.
We thus, assumed the corresponding luminosity of $M_{\rm i} = -14.0$ for all clusters, 
in order to integrate the luminosity function down to the same flux limit. 
To obtain the stellar masses, we adopted different mass-to-light ratio for ellipticals and 
spirals, taken from \citet{kauff03} assuming a \citet{salpeter55} initial mass function (IMF), 
as stated in Equation 21 from \citet{laga08}. The stellar mass estimate
is tied to the choice of IMF. Changing the IMF scales the stellar mass estimate by a fixed factor, 
e.g., from a \citet{kroupa01} IMF to a \citet{salpeter55} IMF with a cut-off at $0.1 M_{\odot}$
results in a factor of 2 increase in the stellar mass \citep[e.g.,][]{kauff03}.
These procedures were also applied in \citet{zlp11}.

When there is a systematic offset between the spectroscopic and photometric magnitudes,
the stellar mass estimate can be systematically biased \citep{kauff03}.
The calibration in \citet{fritz11} shows that the stellar masses computed from 
fiber-aperture magnitudes using DR-7 SDSS photometric data are lower by $\sim$ 0.15 dex
than the values computed from spectroscopic data. Our sample is in similar redshift and 
mass ranges as their sample. We, therefore, corrected this bias in our stellar mass estimates using best-fit
relation between stellar masses computed from the DR-7 SDSS photometric data and from
the spectroscopic data \citep[see the right panel in Fig.~4 in][]{fritz11}. 
All values for stellar masses ($M_{\star}^{\rm corr}$) and optical luminosities ($L_{\rm 500}^{\rm corr}$) 
given in Table~\ref{tab_res} are already corrected and are derived on the basis of $r_{500}$. 
We also give the $\alpha$ and $M^{\ast}$ parameters for the Schechter fit in this table.
A detailed discussion about the impact of the main systematic effects on $M_{\star}^{\rm corr}$ and
$L_{\rm 500}^{\rm corr}$ are presented in Appendix~\ref{opt_systeffec}.

\begin{table*}
\footnotesize
\begin{center}
\caption{Photometric results}
\begin{tabular}{ccccccr@{.}lccc}
\hline 
Name &   \multicolumn{2}{c}{Bright-end} & \multicolumn{2}{c}{Faint-end} & $L_{\rm 500}^{\rm corr} (10^{12} L_{\odot})$ & \multicolumn{2}{c}{$M_{\star}^{\rm corr} (10^{12} M_{\odot})$} & $f_{\star}$& $f_{\rm b}$\\
 &  $\alpha_{1}$ & $M_{1}^{\ast}$ & $\alpha_{2}$ & $M_{2}^{\ast}$ & &  \\ 
\hline
A85	& $-$1.42 $\pm$ 0.09& $-$22.65 $\pm$ 0.32& $-$1.78 $\pm$ 0.080 & $-$ 20.05 $\pm$ 0.51 & 2.32 $\pm$ 0.24 & 6&93 $\pm$ 0.71 & 0.0122 $\pm$ 0.0022 & 0.144 $\pm$ 0.020 \\
A400 	& $-$1.15 $\pm$ 0.06& $-$22.16 $\pm$ 0.19& $-$1.77 $\pm$ 0.002 & $-$ 18.72 $\pm$ 0.31 & 1.61 $\pm$ 0.20 & 4&93 $\pm$ 0.62 & 0.0459 $\pm$ 0.0087 & 0.139 $\pm$ 0.016 \\
IIIZw54 & $-$1.19 $\pm$ 0.08& $-$21.82 $\pm$ 0.65& $-$1.71 $\pm$ 0.044 & $-$ 18.66 $\pm$ 0.64 & 1.38 $\pm$ 0.18 & 3&96 $\pm$ 0.52 & 0.0337 $\pm$ 0.0065 & 0.127 $\pm$ 0.023 \\
A1367 	& $-$1.27 $\pm$ 0.06& $-$22.66 $\pm$ 0.58& $-$1.84 $\pm$ 0.020 & $-$ 18.18 $\pm$ 0.31 & 1.59 $\pm$ 0.27 & 4&75 $\pm$ 0.80 & 0.0225 $\pm$ 0.0050 & 0.129 $\pm$ 0.016 \\
MKW4 	& $-$1.10 $\pm$ 0.16& $-$22.00 $\pm$ 0.36& $-$1.78 $\pm$ 0.020 & $-$ 18.01 $\pm$ 0.13 & 0.31 $\pm$ 0.03 & 0&93 $\pm$ 0.11 & 0.0162 $\pm$ 0.0030 & 0.097 $\pm$ 0.012 \\
ZwCl1215& $-$1.64 $\pm$ 0.03& $-$23.13 $\pm$ 0.31&  	-	       & 	 -	      & 2.71 $\pm$ 0.33 & 8&14 $\pm$ 1.00 & 0.0187 $\pm$ 0.0035 & 0.143 $\pm$ 0.019 \\
A1650 	& $-$1.67 $\pm$ 0.01& $-$23.25 $\pm$ 0.31&	   -	       &     -                & 2.24 $\pm$ 0.38 & 6&67 $\pm$ 1.12 & 0.0156 $\pm$ 0.0034 & 0.141 $\pm$ 0.028 \\
Coma	& $-$1.26 $\pm$ 0.03& $-$21.95 $\pm$ 0.30& $-$1.98 $\pm$ 0.011 & $-$ 18.06 $\pm$ 0.08 & 4.76 $\pm$ 0.54 & 14&00 $\pm$ 1.59 & 0.0226 $\pm$ 0.0041 & 0.154 $\pm$ 0.022 \\
A1795	& $-$1.22 $\pm$ 0.01& $-$21.41 $\pm$ 0.13& $-$1.46 $\pm$ 0.008 & $-$ 19.34 $\pm$ 0.41 & 2.27 $\pm$ 0.34 & 6&99 $\pm$ 1.04 & 0.0157 $\pm$ 0.0032 & 0.141 $\pm$ 0.018 \\
MKW8   	& $-$1.30 $\pm$ 0.11& $-$21.79 $\pm$ 0.58& $-$1.74 $\pm$ 0.009 & $-$ 18.68 $\pm$ 0.70 & 0.61 $\pm$ 0.09 & 1&76 $\pm$ 0.26 & 0.0161 $\pm$ 0.0033 & 0.109 $\pm$ 0.019 \\
A2029   & $-$1.17 $\pm$ 0.07& $-$21.69 $\pm$ 0.17& $-$1.66 $\pm$ 0.020 & $-$ 19.84 $\pm$ 0.12 & 3.18 $\pm$ 0.41 & 9&49 $\pm$ 1.22 & 0.0139 $\pm$ 0.0027 & 0.148 $\pm$ 0.020 \\
A2052	& $-$1.27 $\pm$ 0.10& $-$22.10 $\pm$ 0.48& $-$1.70 $\pm$ 0.011 & $-$ 18.59 $\pm$ 0.09 & 1.37 $\pm$ 0.14 & 4&16 $\pm$ 0.42 & 0.0205 $\pm$ 0.0036 & 0.125 $\pm$ 0.016 \\
MKW3S	& $-$1.36 $\pm$ 0.03& $-$22.14 $\pm$ 0.48& $-$1.57 $\pm$ 0.041 & $-$ 19.12 $\pm$ 0.25 & 3.13 $\pm$ 0.19 & 9&39 $\pm$ 0.80 & 0.0410 $\pm$ 0.0068 & 0.147 $\pm$ 0.017 \\
A2065	& $-$1.10 $\pm$ 0.02& $-$22.00 $\pm$ 0.11& $-$1.62 $\pm$ 0.027 & $-$ 20.08 $\pm$ 0.12 & 2.55 $\pm$ 0.27 & 7&49 $\pm$ 0.57 & 0.0224 $\pm$ 0.0036 & 0.145 $\pm$ 0.042 \\
A2142	& $-$1.00 $\pm$ 0.02& $-$21.84 $\pm$ 0.86& $-$1.68 $\pm$ 0.024 & $-$ 20.56 $\pm$ 0.11 & 3.51 $\pm$ 0.19 & 11&00 $\pm$ 0.96 & 0.0107 $\pm$ 0.0018 & 0.158 $\pm$ 0.023 \\
A2147 	& $-$1.42 $\pm$ 0.05& $-$21.93 $\pm$ 0.12& $-$1.76 $\pm$ 0.015 & $-$ 18.93 $\pm$ 0.12 & 2.24 $\pm$ 0.31 & 6&98 $\pm$ 0.85 & 0.0192 $\pm$ 0.0036 & 0.139 $\pm$ 0.021 \\
A2199 	& $-$1.18 $\pm$ 0.03& $-$21.80 $\pm$ 0.20& $-$1.86 $\pm$ 0.009 & $-$ 18.28 $\pm$ 0.56 & 1.51 $\pm$ 0.27 & 4&72 $\pm$ 0.56 & 0.0179 $\pm$ 0.0033 & 0.128 $\pm$ 0.018 \\
A2255	& $-$1.45 $\pm$ 0.03& $-$23.10 $\pm$ 0.18& 	 -	      &  	 -            & 2.00 $\pm$ 0.26 & 6&18 $\pm$ 0.79 & 0.0151 $\pm$ 0.0029 & 0.141 $\pm$ 0.019 \\
A2589	& $-$1.09 $\pm$ 0.02& $-$21.65 $\pm$ 0.20& $-$1.58 $\pm$ 0.010 & $-$ 18.40 $\pm$ 0.86 & 1.36 $\pm$ 0.14 & 4&19 $\pm$ 0.43 & 0.0223 $\pm$ 0.0039 & 0.125 $\pm$ 0.017 \\
\hline
\end{tabular}
\label{tab_res}
\end{center}
\end{table*}

\section{Results: From stellar mass fraction to the cosmological matter density parameter} 
\label{res}
To fit the trend in our results, we apply the BCES regression fitting method taking into account 
measurement errors in both variables \citep{akritas96}. Through this paper, we apply the BCES regression fitting
taking into account measurement errors in both variables and their covariance \citep{akritas96}.
Confidence intervals correspond to 68\% confidence level.

Figure~\ref{figfrac} shows the stellar, and total baryon fraction as a function of
total mass. We also compared our result to some previous analyses \citep{lms03,giodini09}.
We discuss these results in detail in Section~\ref{sfe}.

In Figure~\ref{mlr}, we show the total mass-to-optical light ratio as a function of total
mass. This result is specifically discussed in Section~\ref{ml}.
In Section~\ref{omegam} we discuss the consequences of determining the
matter-density parameter using galaxy clusters.

\subsection{Star formation efficiency}
\label{sfe}

As clusters of galaxies are the largest 
virialized systems in the Universe, the baryon budget detected
in clusters of galaxies should be relatively representative of
the baryon mass fraction estimate of the Universe,
when considering very massive halos \citep[e.g.,][]{ettori03b,allen08}.

Thus, the total baryon fraction, $f_{\rm b}$, would simply be given 
by $M_{\rm b}=f_{\rm b}~M_{\rm tot}$,
where $M_{\rm b}$ is the sum of $M_{\rm gas} + M_{\star}$, and
$M_{\rm tot}$ is the hydrostatic mass. 
Using our sample of 19 clusters, we investigated 
the dependence of $f_{\star}$, and $f_{\rm b}$ 
as a function of cluster mass.

\begin{figure*}
\centering
\includegraphics[angle=90,width=12cm]{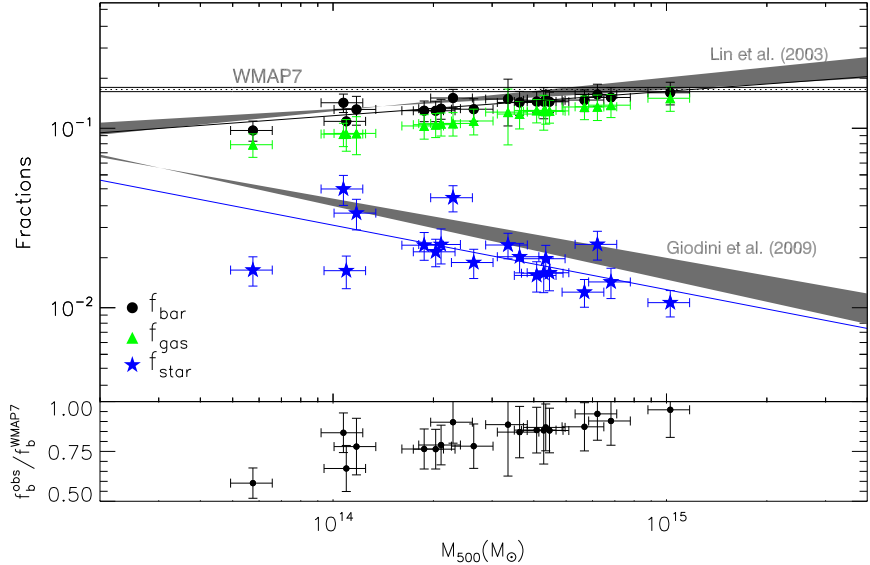}
\caption{\textit{Upper panel}:Total baryon-mass fraction (black circles), gas-mass fraction (green triangles)
and stellar-mass fraction (blue stars) as functions of the total mass. 
The black and the blue solid lines represent power-law fits for the 
total baryon and stellar mass fractions as a function of total mass, respectively. 
The horizontal lines on the top of the plot represent the WMAP-7 result 
\citep[$f_{\rm b}^{\rm WMAP-7} = 0.171 \pm 0.009$][]{jarosik11} with 1 sigma error.
For comparison, we also show the best-fit for the baryon-mass fraction 
from \citet{lms03} and for the stellar mass fraction from \citet{giodini09}.
\textit{Lower panel}: Ration between observational and WMAP-7 baryon mass fraction as 
a function of total mass.}
\label{figfrac}
\end{figure*}

In Figure~\ref{figfrac} we observe a decrease of the stellar
mass fraction, and an opposite trend, although less sharply 
for the gas mass fraction. More massive clusters have gas mass fraction
two times higher and stellar mass fraction around four times lower than less
massive clusters.  
We fit the behavior of the stellar and total baryon-mass fractions
with mass as power laws. Within $r_{500}$ the power-law fit for the stellar-mass fraction
is $f_{\star}= 10^{(-1.54 \pm 0.10)} \times [M_{500}/10^{14} M_{\odot}]^{(-0.36 \pm 0.17)}$,
and the total baryon-mass fraction is given by
$f_{\rm b}=10^{(-0.930 \pm 0.018)}\times [M_{500}/10^{14} M_{\odot}]^{(0.136 \pm 0.028)}$.
At this point, we call the attention to the fact that the 
observed increase of the $f_{\rm b}$ with total mass can be 
partially due to the assumption that the gas mass increases with total 
mass as discussed in Appendix~\ref{mmmg}.

Our study shows a decrease of the stellar mass fraction from
4.5\% to approximately 1.0\%, and taking into account the errors involved, 
one should notice this dependence deviates from a flat distribution within 
$\sim$95\% confidence.
This result suggests that, at 
least on cluster scales, the number of stars formed per unit of halo
mass between low and high-mass clusters are different, that is
the star formation efficiency (SFE) depends on the environment
\citep[as already addressed in several papers ][]{gonz00,lms03,laga08,ettori09,zlp11}.  
In massive clusters, more hot gas and dark matter
is settled in the deep potential wells than the amount of individual
galaxies. In low-mass systems, the accretion of individual
less massive galaxies is more important and more
low entropy gas is brought in to form stars.
Both processes result in lower star formation efficiency for more massive systems.

As a consequence of the dependence of the baryon mass 
fractions on total mass, one does not observe 
a ``flat'' $f_{\rm b}$ distribution as a function of
the total mass on cluster scales. 
As shown in Figure~\ref{figfrac}, the total baryon mass fraction is
lower than the WMAP-7 predicted cosmic fraction,
in line with previous studies, e.g., \citet{ettori03,McCarthy07}, 
but see \citet{Simionescu11}.
In particular, the discrepancy between the observed and the fit value 
becomes larger with decreasing mass 
reaching $\sim 5.5\sigma$ for systems with masses below $3\times
10^{14} M_{\odot}$, and $\sim 3\sigma$ for systems with their
masses higher $3\times 10^{14} M_{\odot}$. 
The disagreement between observations and WMAP result has been 
already pointed out: \citet{gonz07} presented a 3.2$\sigma$ discrepancy, 
\citet{giodini09} found a 5$\sigma$ disparity, and more recently
\citet{andreon10} showed a 6$\sigma$ distinction. 
These studies analyzed cluster and groups sample in a similar mass
range as the one used in this work ($5\times 10^{13}M_{\odot} - 10^{15} M_{\odot}$). 
\citet{giodini09} analyzed 91 objects spaning a range in 
$M_{500}$ of $\sim10^{13}-10^{15} M_{\odot}$. 
\citet{andreon10} investigated 52 clusters and groups 
in the same mass range as the latter authors, and
\citet{gonz07} considered 24 galaxy clusters in a broader interval of mass, ranging from
$M_{500} = 10^{12} M_{\odot}$ up to $M_{500} \sim 8\times10^{14} M_{\odot}$.

It is worth noting that the most massive 
cluster in our sample has a total mass of $M \sim 10^{15} M_{\odot}$ 
and its baryon mass fraction already reaches $\sim$96\% of the cosmic value. 
Extrapolating our fit for the total baryon mass fraction, $f_{b}$ reaches the WMAP-7 
prediction for clusters with $M \ge 1.6 \times 10^{15} M_{\odot}$. We show the ratio
between observed and  WMAP-7 values for $f_{\rm b}$ as a function of total mass in the
lower panel of Fig.~\ref{figfrac}. Hydrodynamical simulations claim that baryons in 
clusters have undergone a large depletion during the formation of these structures.

Although \citet{white93} claim that within a sufficiently large radius the mean
baryon fraction must take the global value, \citet{frenk99} assert that
the depletion parameter, that is the ratio between the observed and the predicted 
baryon mass fraction is $\Upsilon=0.925$, which is in line with more recebt estimate
\citep[e.g.,][]{ettori06,young11}.
This means that even considering a large radius, e.g. $r_{200}$, some baryons are missing 
also in massive clusters.

The comparison of the observed baryon budget to the WMAP-7 cosmic fraction \citep{jarosik11}
consistently suggests that the amount of missing baryons increases
toward low-mass systems. Recently, \citet{dai10} proposed that the baryon 
loss mechanism is primarily controlled by 
the depth of the potential well of the system.
For deep potential wells, such as rich clusters, baryon loss is not significant, 
however, for lower-mass clusters, baryon loss becomes
increasingly important. Such a mechanism
could be the pre-heating of baryons before they
collapse. In this picture, the baryons never fall into groups
and low-mass clusters, but remain well beyond $r_{200}$. 
Alternatively, the gas falls
into the potential-wells and is subsequently removed by feedback
provided from SNe and AGNs, which should also play a major role 
for low-mass systems.

Some missing baryons are suggested, which can partially account for
the discrepancy between the observed and the
predicted cosmic baryon mass fraction. 
Intra-cluster light (ICL) is one of the most
important missing baryons. 
Observational results have shown that the ICL can account for 6\%--22\% of 
the total cluster light in the $r$-band 
\citep[e.g.,][]{kb07,gonz07,pierini08}, and being 
an order of magnitude lower, it would account for $\sim 2\%$
in the total baryon mass fraction. 
Recent numerical simulations performed by \citet{puchwein10} showed that 
the amount of gas and ICL removed by active galactic nucleus (AGN) heating from the central 
regions of clusters and driven outwards ($r > r_{500}$) depends on cluster mass, being 
higher in low-mass systems. 
\citet{puchwein10} estimate that for a cluster with total mass $M=4 \times 10^{13} M_{\odot}$
the ICL account for 58\% of the total stars. 
Considering this latter result, there is no significant missing baryons 
as suggested by e.g. \citet{afshordi07}. 
Just the ICL and some part of the ICG would be sufficient to explain the
difference between observations and cosmological prediction. 
However, such high amount of ICL would be visible in current observations.

Another reasonable possibility is a dependence of the baryon budget
as a function of radius \citep[e.g.,][]{henri94,ettori99}.
It might be that in the limiting radius of X-ray observations ($r_{500}$) the
total baryon budget is still underestimated. 
Although \citet{allen02} show that the gas mass fractions in the clusters 
asymptote towards an approximately constant value at a radius $r_{2500}$,
most of the numerical simulation results and observational data show an
evident increase of the baryonic content with cluster radius.
\citet{Vikh06} showed that the gas mass fraction increases with radius as a power 
law of overdensity. Recently, \citet{Simionescu11} showed that the baryon mass fraction 
increases dramatically with cluster radius for the Perseus cluster. 
This cluster is one of the few clusters having deepest X-ray data to 
measure the baryon mass fraction up to the virial radii. They found that, for the Perseus cluster,
the gas mass fraction alone (without considering the stars) reaches the baryonic cosmic mean value
at about half of $r_{200}$ and at $r_{200}$ it is $\sim38\%$ above the mean baryonic value.
However, it is important to have in mind that this is an isolate result and it is important to
analyze the dependence of $f_{\rm gas}$ (determined within the virial radius) on total mass
for a sample of clusters. For instance, \citet{ReipPHD03} 
studied the gas mass fraction within $r_{200}$ for 106 clusters, 
and Perseus is among the five clusters with the highest $f_{\rm gas}$.
This study can be readdressed using Suzaku data.

\subsection{Total mass-to-optical light ratio}
\label{ml}

In this work we used independent X-ray and optical methods to compute the total 
mass-to-optical light ratio for our sample, and both optical luminosities and total masses were
homogeneously determined. Compared to other estimates, the determination of total masses using X-ray is a robust 
and a preferable method since it involves lower uncertainties
\citep{nagai07,meneguetti10}.

For $\Lambda$CDM cosmological model, the mass-to-light ratio ($M/L$)  
in high-mass systems is supposed to be approximately independent of halo mass. 
It has been generally found that $M/L$ increases
with halo mass but there is an approximate plateau 
in $M/L$ values for the richest bound systems \citep[e.g.,][]{david95,cirimele97,sheldon09}.
The existence of this plateau is sometimes 
taken as evidence that the measured values of $M/L$ do indeed 
represent the universal value.

However, recently, most authors have found a dependence between the
$M/L$ and the total mass, even for massive clusters of galaxies.
Assuming a power-law relation, $M_{500}/L_{\star} \propto
M_{500}^{~\alpha}$, those authors have found $\alpha$ to be in the
range of 0.2--0.4, in both optical and near-infrared bands, over a
large mass range \citep{adami98,bahcall02,rines04}.

\begin{figure}
\centering
\includegraphics[angle=-90,width=8.3cm]{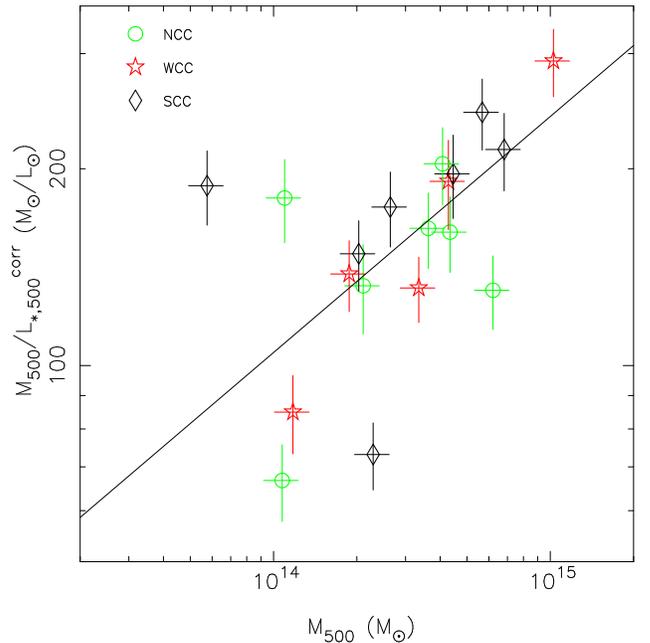}
\caption{Total mass-to-light ratio as a function of total mass of the system.
The black, red and green colors denote strong, weak cool-core and non cool-core clusters, respectively.}
\label{mlr}
\end{figure}

In Figure~\ref{mlr} we show the dependence of the total mass-to-optical light ratio 
($M_{500}/L_{\star}$) on total mass.
The best-fit for the mass-to-optical light ratio follows
$M_{500}/L_{\star} = 10^{(2.02 \pm 0.10)} 
\times [M_{500}/10^{14} M_{\odot}]^{(0.361 \pm 0.169)}$. 
Considering the sample analysed in this work, a plateau is not clearly present 
for the massive end, and the best-fit
leads to a $M_{500}/L_{\star} = 241 M_{\odot}/L_{\odot}$ for clusters
with total mass of $10^ {15}M_{\odot}$. 

We see from Figure ~\ref{mlr} that non cool-core clusters present
a higher dispersion when compared to cool-core clusters.
This may indicates that the dynamical state of the cluster
should be considered when analyzing the mass-to-light ratio, because
it might follow that for a large and homogeneous sample of cool-core clusters
a plateau becomes more evident. 
Thus, in order to better investigate
whether there is a plateau at the massive end, 
further studies should consider a 
complete flux-limited sample of cool-core clusters.

The dependence of $M_{500}/L_{\star}$ shown in Figure~\ref{mlr} with cluster mass
is a direct consequence of the varying star formation efficiency
in galaxy clusters. Since low-mass systems have higher star formation
efficiency compared to a massive cluster, the optical light will
not vary uniformly and thus, the total mass-to-optical light 
shows a dependence on cluster mass. 

\subsection{The cosmological parameter $\Omega_{\rm m}$}
\label{omegam}

The results presented in the Sections \ref{sfe} and \ref{ml}
have important implications on the cosmological
use of galaxy clusters. Using the \citet{oort58} technique, 
one can determine the matter density parameter following
$\Omega_{\rm m}=(M/L) \times \rho_{\rm L}/\rho_{\rm c}$, where $\rho_{\rm c}$ is the
critical density, and $\rho_{\rm L}$ is the typical luminosity density
of the universe. 
A fundamental assumption in
this application is that $M/L$ of clusters is a
fairly representative measurement of the universal value. 
However, the mass-to-light ratio has been found to increase with cluster
mass. Assuming the mass-to-light ratio range shown in Figure~\ref{mlr} 
(from 70 $M_{\odot}/L_{\odot}$ up to 300 $M_{\odot}/L_{\odot}$), and
adopting $j_{i} \sim 2.12 \times 10^{8} h_{100}~L_{\odot}/\rm Mpc^{3}$ for the luminosity density 
\citep[see Table 3 from][] {blanton03}, 
we have the total matter density parameter 
spanning the range $0.07 < \Omega_{\rm m} < 0.27$.
Taking into account the errors involved in this method to derive
both total mass ($\sim10\%$) and the optical light 
(a mean value of about $20\%$), 
we can determine $\Omega_{\rm m}$ within $\sim14\%$ of uncertainty.
It should be noted that since we did not reach the claimed
plateau in mass-to-light ratio towards the massive end, the upper limit
obtained here for the matter density parameter ($\Omega_{\rm m} = 0.3$),  
although consistent with the WMAP-7 concordance value within the 
errors \citep{jarosik11}, could still be an underestimate of $\Omega_{\rm m}$. 

Another way to determine the matter density parameter is considering that 
the ratio of baryonic-to-total mass in clusters should closely match the cosmological 
parameters $\Omega_{\rm b}/\Omega_{\rm m}$. 
Thus, adopting $\Omega_{b}h^{2} = 0.02260 \pm 0.00053$ \citep{jarosik11}, 
and the total baryon fraction computed in this work (and listed in Table~\ref{tab_res})
we obtained $ 0.15 < \Omega_{\rm m} < 0.30$,
a narrower interval compared to the previous method.
This shows that the stellar and gas mass estimates from optical
and X-ray data provide better determination of $\Omega_{\rm m}$
when compared to the use of $M_{500}/L_{\star}$.
The first advantage is that to derive the gas mass one do not assume hydrostatic equilibrium,
and thus the dynamical state of the cluster will not bias this determination.
Then, the uncertainties involved in this determination 
are around 5\%-10\% \citep{nagai07,meneguetti10}.
Also, X-ray gives a 3D view of the cluster when deprojection is done, while total
light is always derived from a 2D projection of galaxies in the CMD.
There are also the difficulties already pointed out in the Section~\ref{intro} and
discussed in \citet{girardi00} related to the luminosity estimates. 
Moreover, there  has been some concern with the $M/L$ 
method because recent numerical simulations
suggest that light is a biased tracer of dark matter, providing a biased value for 
 $\Omega_{\rm m}$ \citep{ostriker03}. 
The Oort method \citet{oort58} necessarily focus on denser regions of the galaxy distribution, and thus 
the estimate of $\Omega_{\rm m}$ relies on the assumption that the galaxy 
population in these regions is representative of the universe as a whole.
However, it is well-known that the galaxy population in clusters
differs from the field population since the work from \citet{dressler80}.
Since rich clusters of galaxies encompasses 
large regions (of several Mpc), observations that extend up to the 
virial radius ($r \sim r_{200}$) should, in principle, contain a sufficient
collapsed region and provide a fair mass-to-light ratio. 
Nevertheless, it is important to stress that in this 
work we just examine a cluster volume of radius $r_{500}$,
what can also affect our estimates.   

It is worth mentioning the effect of the ICL on both determinations.
In both cases, the ICL would contribute to lower the $\Omega_{\rm m}$ estimates. 
Using the \citet{oort58} technique, the ICL would account for the total light
and then this ratio would be smaller than the ones obtained here. 
Thus, if on the one hand the upper limit for $\Omega_{m}$ could still be higher
when reaching the plateau for massive clusters; on the other hand, the 
inclusion of the ICL can contribute to a lower estimate of the matter density
parameter.
Furthermore, using the total baryon-to-mass fraction,
the matter density parameter is given by 
$\Omega_{\rm m} \sim \Omega_{\rm b} \times M_{\rm tot}/M_{\rm b}$, and
the ICL would account for the total baryon estimates ($M_{\rm b}$). 
Consequently one would obtain a lower determination of the matter density parameter.
Thus, in both cases, the derived values in this work are biased up 
for not considering the ICL contribution. 


\section{Conclusions} 
\label{conc}

We studied a representative sample of 19 clusters, and investigated this
sample as a robust local reference to constrain the baryon mass
fraction and its mass dependence on cluster scales for future
studies. Our conclusions are as follows.

\begin{itemize}
\renewcommand{\labelitemiv}{$-$}

\item We obtained optical luminosities and total masses for a 
sample of 19 clusters in a homogeneous way. Thus, we computed
total mass-to-optical light ratio using two independent methods, what
present a considerable advantage over previous studies.

\item The total mass-to-optical light ratio decreases toward low-mass systems, following the relation
$M_{500}/L_{\star} = 10^{(2.02 \pm 0.10)} \times [M_{500}/10^{14} M_{\odot}]^{(0.361 \pm 0.169)}$.
This result is a direct consequence of the varying SFE in cluster scales. 
In this work the $M_{500}/L_{\star}$ 
varies from $\sim$60 $M_{\odot}/L_{\odot}$ up to almost 300 $M_{\odot}/L_{\odot}$.
Total mass-to-optical light ratio does not show evidence of a flattening,
and the best-fit leads to a $M_{500}/L_{\star} = 241 M_{\odot}/L_{\odot}$ for clusters
with total mass of $10^ {15}M_{\odot}$. 
Our results indicate that this flattening towards massive clusters
should be more evident when considering a flux-limited sample of 
cool-core clusters. We observed that non cool-core clusters
present a large dispersion when compared to more relaxed systems
and may introduce some bias in the final dependence. 

\item Both $f_{\rm b}$ and $f_{\star}$ show a mass
dependence. Within $r_{500}$, the power-law fits are 
$f_{\star} = 10^{(-1.54 \pm 0.10)} \times [M_{500}/10^{14} M_{\odot}]^{(-0.359 \pm 0.170)}$,
and $f_{\rm b} = 10^{(-0.93 \pm 0.018)} \times [M_{500}/10^{14} M_{\odot}]^{(0.136 \pm 0.028)}$ 
for the stellar, and total baryon mass fraction, respectively. 
The observed increase of the $f_{\rm b}$ with total mass can be 
partially due to the increase of the gas mass 
with total mass as discussed in Appendix~\ref{mmmg}.
Toward the high-mass end, the value gradually approaches the
WMAP-7 prediction, reaching it at $M_{500} = 1.6 \times 10^{15} M_{\odot}$,
when extrapolating our fit for the total baryon mass fraction.
However, the most massive cluster in our sample has a total mass 
of $M_{500}=1.026 \times 10^{15} M_{\odot}$ and 
its baryon mass fraction already reaches $\sim$96\% of the cosmic value.
The amount of missing baryons with respect to the
WMAP-7 predicted fraction increases toward lower systems.
This can be compensated by ICL and a significant part of the gas that
is driven outwards ($r > r_{500}$) due to AGN feedback, especially in
lower mass-systems.

\item 
The SFE is lower in more 
massive clusters, increasing towards the 
low-mass end. The rapid decrease observed  in the
stellar mass fraction as a function of total mass 
suggests a significant change in the efficiency
of star formation with cluster mass 
\citep[as already addressed in several papers:][]
{gonz00,lms03,laga08,ettori09}.

\item We derived the matter density parameter using the \citet{oort58} technique
and also from the baryon-to-total mass ratio. Using these two approaches, 
we obtained $0.07 < \Omega_{\rm m} < 0.3$ 
and $0.15 < \Omega_{\rm m} < 0.27$, respectively. 
Using the baryon-to-total mass ratio to compute $\Omega_{\rm m}$
seems to give narrower range and more accurate values.
Since we did not consider the ICL contribution, the obtained values
for the matter-density parameter are possibly biased low. 
\end{itemize}

\begin{acknowledgments}
\small
The XMM-\textit{Newton} project is an ESA Science Mission with instruments and contributions
directly funded by ESA Member States and the USA (NASA). The XMM-\textit{Newton} project
is supported by the Bundesministerium f\"ur Wirtschaft und Technologie/Deutsches Zentrum f\"ur Luft- und
Raumfahrt (BMWI/DLR, FKZ 50 OX 0001) and the Max-Planck Society.
We thank the referee, Stefano Ettori, for relevant questions and fruitfuil discussions that
improved the quality of this manuscript.
T.~F.~L acknowledges G.~B.~Lima Neto and F.~Durret for constructive discussions.
T.~F.~L also thanks the financial support from FAPESP (grants:  2006/56213-9, 2008/04318-7)
and CAPES (grant: BEX3405-10-9). We thank Jacopo Fritz for providing
the best-fit relation between stellar masses computed from the DR-7 SDSS photometric and
spectroscopic data. Y.~Y.~Z also acknowledges support from the German BMBF through the 
Verbundforschung under grant 50~OR~1005. T. H. R. acknowledges support by 
the DFG through Heisenberg grant 1462/5. This work was supported by the Deutsche 
Forschungsgemeinschaft under the Collaborative Research Center TR-33.

\end{acknowledgments}

\bibliography{refs2}

\appendix

\section{$M_{\rm gas}$-$M_{500}$ relation}
\label{mmmg}
Since total mass estimates are derived under the assumption of hydrostatic equilibrium (EQ), 
and the present sample contains clusters in a wide variety of dynamical states 
(here separated in strong cool-core, 
weak cool-core and non cool-core clusters), the EQ
assumption may not be valid in all cases. Thus, we preferred to 
derive total masses using the scaling relation. 
Recent observational investigations using a variety of cluster samples have demonstrated 
that the gas mass is indeed a low-scatter mass proxy \citep{Maughan07,Arnaud07}. 
To construct a scaling-relation between total and gas mass, 
as shown in Fig.~\ref{figmgmt}, we used 41 
dynamically relaxed clusters from \citet{Vik06}, \citet{Arnaud07}, 
\citet{Bohringer07}, and \citet{Sun09}, where the dynamical equilibrium assumption 
can be fairly applied to derive total mass. 
Then, using $M_{\rm tot}-M_{\rm gas}$ relation stated in Eq.~\ref{mgmtot}, 
we can recover the total mass of our sample. 
In this way, for the non-relaxed clusters, the total mass can be quite accurately determined
\citep[$\sim$10\%,][]{nagai07,Zhang08}. Also, according to numerical simulations \citep[e.g.,][]{nagai07}, 
the total ICM mass is measured quite accurately ($<$6\%) in all clusters, 
irrespective to the dynamical state.

We also should highlight that the way the total mass was computed from the 
scaling relation, $\log M_{\rm tot}=\rm A + B\times \log M_{\rm gas}$,
one obtaines $M_{\rm tot}=10^{\rm A} \times M_{\rm gas}^{\rm B}$, and 
$f_{\rm gas} = M_{\rm gas}/M_{\rm tot} = 10^{-A/B}\times (M_{\rm tot})^{(1-B)/B}$. Thus, 
being $f_{\rm gas} \propto (M_{\rm tot})^{(1-B)/B}$, when $B$ is positive and lower than 1, 
the gas mass fraction increases with total mass. Since we obtained $B=0.827$, we assume
$f_{\rm gas} \propto M_{\rm tot}^{0.2}$.
\setcounter{figure}{0}

\begin{figure}[h!]
\centering
\includegraphics[angle=-90,width=10cm]{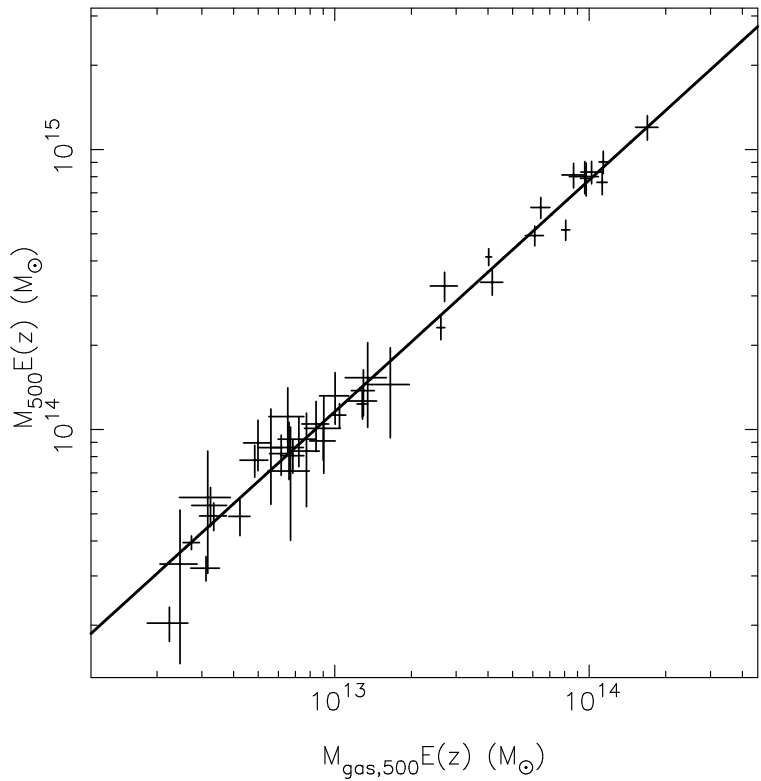}
\caption{Total mass as a function of gas mass for 41 groups and clusters.
The continuous line is the power-law fit for the data.}
\label{figmgmt}
\end{figure}

\renewcommand{\thefigure}{\thesection.\arabic{figure}}
\section{Color-Magnitude diagrams and the luminosity function fits}
\label{ap}

The elliptical galaxies in individual clusters form a red-sequence (RS) with a well-defined slope and 
small scatter \citep{bower92,bower92b}. Specifically, the existence of the red-sequence at higher 
redshifts indicates that cluster elliptical galaxies are a passively evolving population in which the
reddening of massive galaxies is the result of a mass-metallicity relation rather than an age effect
\citep{kodama97,kauffmann98}. From the observational point-of-view, studies 
have shown that for nearby clusters ($z < 0.05$) the slope of 
the red-sequence is almost constant \citep{gladders98,romeo08}. Since we have almost
half of the clusters in our sample with redshifts higher than the above mentioned redshift, 
we did not fix the
slope of the RS but left it to vary. We show in Figure~\ref{cmr} the red-sequence for all clusters  in our sample.
Even not fixing the slope, we obtained well-defined red-sequence in which the mean slope is 
$b=-0.036 \pm 0.010$. 

For all galaxies belonging to the cluster, we build the luminosity function.
The luminosity function is the distribution of all morphological types
of cluster galaxies over the magnitude. This distribution is generally fit
by the Schechter function \citep{schech76}. In Figure ~\ref{cmr}, we present
the luminosity function fits for all nineteen clusters. 
In particular, the slope of the faint end of the LF is a direct indicator
of the importance of dwarf galaxies, which are expected to
be more fragile in the environment of clusters.
The great majority of studies of the LF indicate faint-end
slopes in the range of $-0.9$ to $-1.5$, but these mostly did not reach
very faint magnitudes \citep[see Table 1 in][and references therein]{dePropris03}. 
Recent studies using deep imaging has shown 
deep estimations of the faint-end slope within the range $-2.29 < \alpha < -1.07$ 
\citep[see Table A1 from][and references therein]{boue08}. 
Our mean values for the slope of the bright population and of the faint-end are
$<\alpha_{1}> = -1.28 \pm 0.18$ and $<\alpha_{2}> = -1.72 \pm 0.13$, respectively.
For the characteristic magnitude we obtained $M_{1}^{\ast}=-22.16 \pm 0.50$ and 
$M_{2}^{\ast}=-18.97 \pm 0.85$ for the bright- and for the faint-end, respectively. 

\setcounter{figure}{0}
\begin{figure}[ht!]
\centering
 \includegraphics[width=0.8\textwidth]{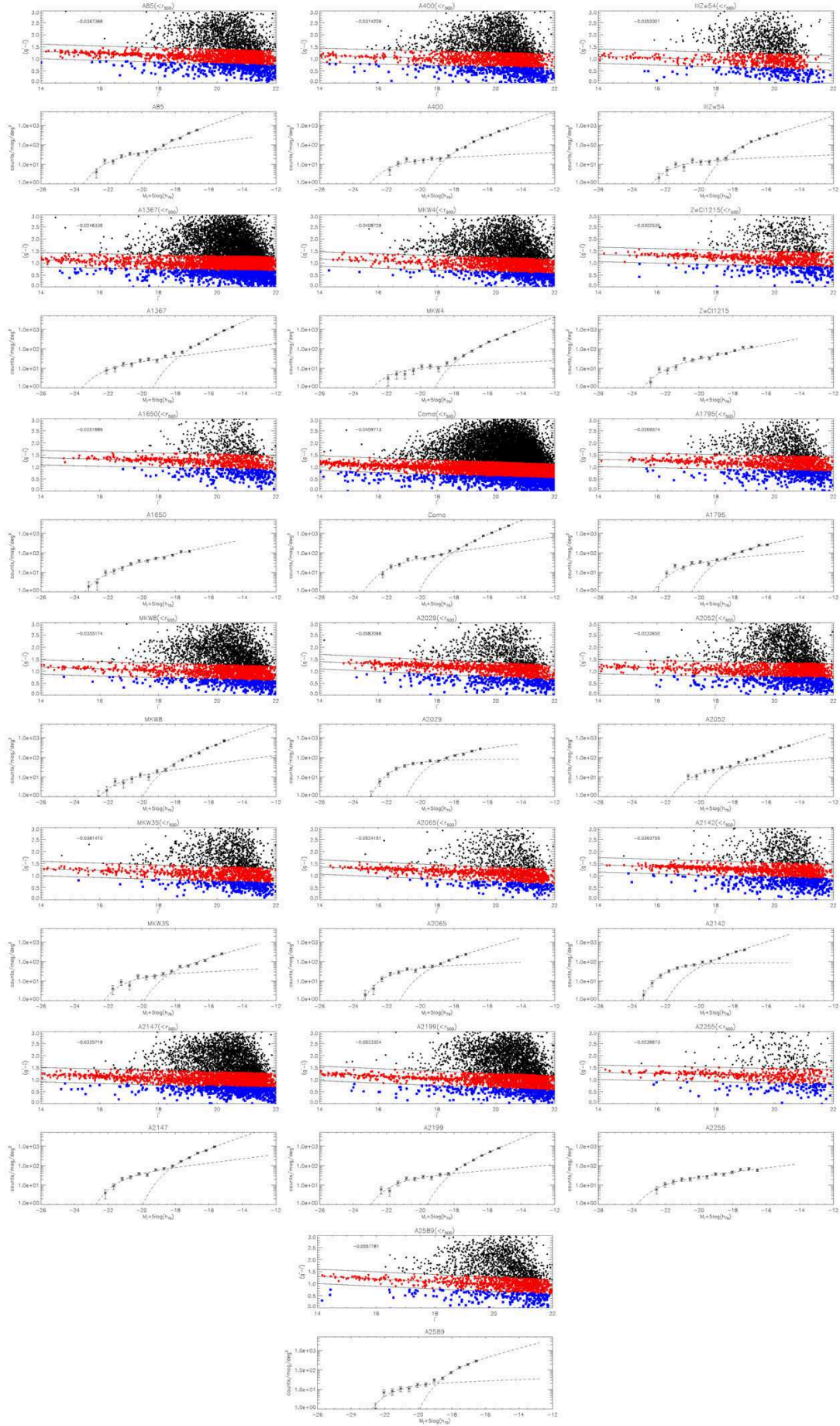}
\caption{Color-magnitude diagrams and luminosity function fits for our sample.}
\label{cmr}
\end{figure}

\section{Impact of systematic uncertainties in the photometric results}
\label{opt_systeffec}

One of the basic observational result of the present study is the stellar masses estimate.
To correct determine the stellar mass we followed the description in Sec.~\ref{photodata},
and here we will discuss the principal steps that account for the systematic uncertainties involved 
in the total optical luminosity and stellar mass estimates. These systematic uncertainties are
summarized in Table~\ref{tab_sisteffec}.
First, we can mention that the stellar masses derived from photometric DR7 data can be 
systematic biased low due to the offset between photometric and spectroscopic magnitudes.
This bias can account for $\sim10\%$ of uncertainty \citep{fritz11}.
Then, we can highlight the impact of assuming one Schechter function to describe the
overall distribution when one have a steep population faintward. This can affect the
determination of $M^{\ast}$ and $\alpha$ parameters. When not using two functions, 
the determination of these parameters can systematic bias up the total luminosity 
determination by $6-15\%$, what will be directly translated to the $M^{\star}$ estimates.
Also related to the Schechter fit, we can adress the systematics of the covariance 
between $M^{\ast}$ and $\alpha$ parameters. If not taken into account, the uncertainties 
related to these parameters can biased low the total luminosity estimates by about 5\%.
The major systematic effect may be introduced by the adopted IMF when 
assuming a mass-to-ligh ratio to obtaine the stellar mass. For instance,
a change from a standard \citet{salpeter55} to \citet{kroupa01} IMF increses
the $M/L$ by a factor of two. This translates into an increase by the same amonut 
in the stellar mass in our systems. 
Finally we can adress the ICL contribution. In this work, we did not consider the total
light contribution to the stellar mass estimates. Thus, our values can be 2\% 
systematic biased low (see Sect.~\ref{sfe}).

\begin{table}[h!]
\small
\begin{center}
\caption{Main systematic uncertainties in stellar mass determination}
\begin{tabular}{cc}
\hline 
uncertainty & \% (approx.)\\
\hline
photometric magnitude & -10 \\
Scheschter fit & + 6-15 \\
covariance between $M^{\ast}$ and $\alpha$ &  -5\\
IMF & +100 \\
not considering ICL & -2\\
\hline
\end{tabular}
\label{tab_sisteffec}
\end{center}
\end{table}

\end{document}